\documentclass{mem}
\usepackage{natbib}\usepackage{txfonts}\usepackage{balance}
\usepackage{graphicx}
\usepackage[a4paper]{hyperref}
\idline{75}{282}
\begin{document}
\def\teff{$T\rm_{eff }$}
\def\kms{$\mathrm {km s}^{-1}$}

\title{Observing the strong gravity regime of\\
       accreting black holes with Simbol-X}

   \subtitle{}

\author{R.~W. Goosmann\inst{1,2}, M. Dov{\v c}iak\inst{1},
\and V. Karas\inst{1}}

  \offprints{R. W. Goosmann}
 
\institute{
Astronomical Institute, Academy of Sciences,
Bo{\v c}n\'i II 1401a,
CZ--14131 Praha 4,\\
Czech Republic,
\email{goosmann@astro.cas.cz}
\and
Observatoire de Paris, Section de Meudon,
5 pl. Jules Janssen,
F--92190 Meudon,
France
}

\authorrunning{Goosmann, Dov{\v c}iak, \& Karas}

\titlerunning{Strong gravity regime of accreting black holes}

\abstract{The X-ray reflection features of irradiated accretion disks around
  black holes enable us to probe the effects of strong gravity. We investigate
  to which precision the reflection signs, i.e. the iron K-line and
  the Comptonized hump, can be observed with Simbol-X for nearby Seyfert
  galaxies. The simulations presented include accurate computations of the
  local reprocessed spectra and modifications due to general relativistic
  effects in the vicinity of the black hole. We discuss the impact of  global
  black hole parameters and of the irradiation pattern of the disk on the
  resulting spectra as they will be detected by the Simbol-X mission.
    }

\maketitle{}

The X-ray spectra of many accreting black holes are known to reveal
relativistically modified reflection features, notably a fluorescent iron
K-line complex \citep[for a review see][]{reynolds2003} and a Comptonized
hump \citep[e.g.][]{lightman1988}. With its broad photon energy coverage the
Simbol-X satellite is particularly suited to observe such X-ray
reflection spectra. In this proceedings note, we show that Simbol-X will
be able to distinguish X-rays being re-emitted at different radii of the
innermost accretion disk.   

We assume that the X-rays are produced by magnetic flares co-orbiting with the
disk so that the reflection features originate in underlying orbiting
spots. Detailed reflection spectra of such spots are modeled using the coupled
radiative transfer codes Titan and Noar \citep{dumont2000,
dumont2003} and the relativistic effects are added applying the ray-tracing
model KY \citep{dovciak2004} inside XSPEC. Our modeling procedure
is described in detail in \citet{goosmann2007}. We assume a black hole with $M
= 3 \times 10^7 \, M_\odot$ that is maximally spinning and that carries an accretion
disk being in hydrostatic balance (before the flare goes off). We consider
orbiting spots at the disk radii $R = 4 \, R_{\rm g}$ and $R = 21 \, R_{\rm
  g}$ ($R_{\rm g} = GM/c^2$). Simulated data are obtained for the ``observed''
spot emission at both radii using a modified version of the kyl1cr model
in XSPEC and the currently available response matrices for the MPD
and CZD detectors of Simbol-X. The observation time is set to one
Keplerian orbit in both cases and the flux rates are adjusted to an XMM-Newton
observation of the Seyfert galaxy NGC~3516 \citep{iwasawa2004}. 

\begin{figure*}[t!]
  \centering
  \resizebox{0.9\hsize}{!}{\includegraphics[clip=true]{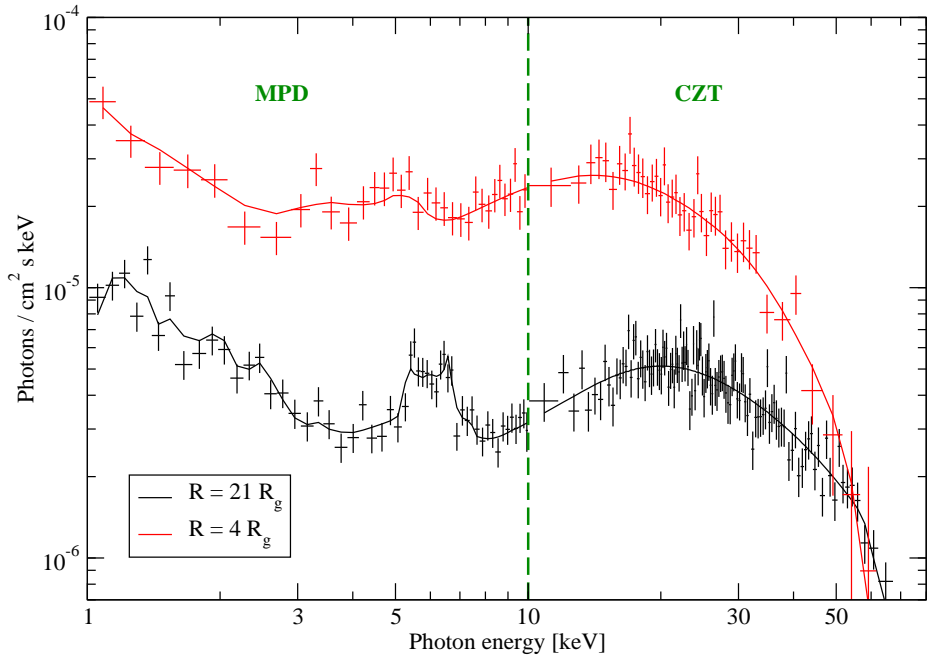}}
  \caption{\footnotesize Simulated MPD and CZD spectra of an
    orbiting flare spot at $R = 4 \, R_{\rm g}$ (top) and $R = 21 \, R_{\rm
    g}$ (bottom). The underlying XSPEC models are shown as solid
    lines. The data sets contain a vertical offset for clarity.}  \label{fig}
\end{figure*}

In Fig.~\ref{fig} the time-integrated spectra for both orbiting flares are
plotted over the spectral range covered by the two detectors. While the
overall spectral shape is similar for spots at both disk radii, the position
and the shape of the reflection features differ significantly. 

At $R = 4 \, R_{\rm g}$ the impact of the gravitational redshift pushes the
maximum of the Comptonized hump below 15 keV and the strong projected velocity
gradient along the orbit almost entirely flattens the profile of the iron line
complex. At $R = 21 \, R_{\rm g}$, on the other hand, the Compton hump is
centered on 20~keV and the iron line profile is clearly visible showing a
double horn. The data further suggests the appearance of separate soft X-ray
emission lines below 3 keV.    

For rapidly spinning black holes, it might thus be impossible to detect
reprocessed emission from the innermost accretion disk (below $R = 6 \,
R_{\rm g}$) in the iron line band, but the position of the Comptonized hump
can give an alternative handle on the location of the reprocessing site. For
larger disk radii, the fact that Simbol-X observes both reflection
features simultaneously puts important constraints on the emission radius and
also on the applied reprocessing models.  

\begin{acknowledgements}
We thank the organizers of the Simbol-X workshop for an interesting conference. 
\end{acknowledgements}

\bibliographystyle{aa}

\end{document}